# Comparison of UHE Composition Measurements by Fly's Eye, HiRes-prototype/MIA and Stereo HiRes Experiments

P. Sokolsky[a], John Belz[a,b] and the HiRes Collaboration
*(a) High Energy Astrophysics Institute, University of Utah, Salt Lake City, Utah 84112, USA*
*(b) University of Montana, Missoula, Montana 59812, USA*
Presenter: P. Sokolsky (ps@cosmic.utah.edu), usa-sokolsky-P-abs1-he14-oral

We compare the elongation rate and Xmax distributions for three air-fluorescence experiments: Stereo Fly's Eye, HiRes/MIA, and HiRes. A shift of 13 gm/cm$^2$ of the stereo Fly's Eye data, well within the quoted systematic errors, brings the elongation rates and the Xmax distributions of all three experiments into reasonable agreement. We explore the implications of this combined data set ranging from $10^{17}$ eV to near $10^{20}$ eV.

## 1. Introduction

There have now been three fluorescence-based experiments which have reported results on the cosmic ray composition using the shower maximum (Xmax) method. These are the Stereo Fly's Eye [1], the HiRes Prototype/MIA[2] and the HiRes[3] experiments. The Fly's Eye collaboration reported on data taken with the Stereo Fly's Eye configuration. In this experiment, two fluorescence detectors were positioned 3.5 km apart. The effective energy range extended from $10^{17}$ eV to $3\times10^{19}$ eV. This experiment utilized optics that resulted in a 5×5 degree pixel size, and the geometry of the event was determined by the intersection of the two event-detector planes, with resulting resolution in Xmax of about 45 gm/cm$^2$. Both Fly's Eye detectors had mirrors that covered an elevation angle range from 0 to 90 degrees, though FE II had only partial azimuth coverage.

The HiRes Prototype/MIA experiment was a hybrid fluorescence - muon array detector. The fluorescence detector, a fourteen mirror prototype of the current HiRes detector had a 1×1 degree pixel size and was located 3.5 km from the center of the MIA muon array. The muon array sampled a part of muon lateral distribution. The effective area of this "array" was thus set by the muon lateral distribution function and was roughly a 3 km radius centered on MIA. The smaller pixel size of the HiRes detector allowed triggering to lower energies (approaching $5\times10^{16}$ eV ) while the effective area of MIA limited the energy reach to just above $10^{18}$ eV. The Xmax resolution was 45 gm/cm$^2$. The geometry of the event was determined by using the HiRes event-detector plane and timing information from the fluorescence detector and the MIA ground array.

The HiRes stereo detector consists of two fluorescence detectors separated by 12 km. With 1×1 deg pixel size, the Xmax resolution is 30 gm/cm$^2$. The large separation between detectors and the limited elevation angle range (3 to 32 degrees) limits the energy range to > $10^{18}$ eV with the statistical reach approaching $10^{20}$ eV.

## 2. Previous Results

All three experiments have published results on the cosmic ray composition in various energy ranges. Figure 1 shows the elongation rate results, together with expectations from various hadronic models for a purely protonic and a purely Fe composition. In each case, there is evidence for a transition from a predominantly heavy composition, near $10^{17}$ eV to a much lighter composition at higher energies. While the qualitative picture is consistent between the experiments, the details appear to be different, with the Stereo Fly's Eye results indicating a more gradual transition while the combined HiRes Prototype/MIA and HiRes results indicate a transition to a light composition occurring more abruptly near $10^{18}$ eV. The fits to the elongation rates are also somewhat different, with the Stereo Fly's Eye quoting an overall elongation rate of 78.9 ± 3.0 gm/cm2/decade, while the HiRes Prototype/MIA experiment quotes 93.0 ± 8.5 gm/cm2/decade below $10^{18}$ eV and 54.5 ± 6.5 gm/cm$^2$/dec above this value.

The elongation rate figures only show the error on the mean of the Xmax distribution in each energy bin. All three experiments quote systematic errors between 20 to 25 gm/cm$^2$ on the absolute position of Xmax. These are largely dominated by mirror Čerenkov light subtraction. Differences in elongation rate slopes between different experiments can in principle also be produced by non-linear energy response. The Fly's Eye Stereo experiment shows a significant ankle structure with a minimum near $3\times10^{18}$ eV. The HiRes detector sees a very similar ankle structure with a minimum at the same energy. The flux normalization between the two spectra is also similar, reinforcing the evidence that the energy scales between the two experiments are very similar, at least in this energy range.

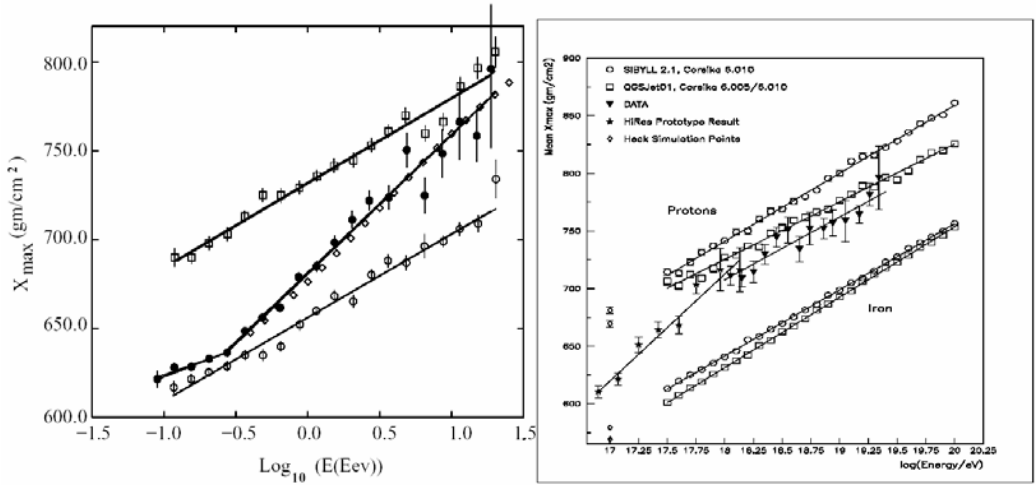

**Figure 1.** Elongation rate for Stereo Fly's Eye (left) and HiRes/MIA and HiRes (right) experiments

## 3. Applying a Systematic Shift

The resolution in Xmax tends to improve with increasing energy as more of the extensive shower is measured. Čerenkov subtraction effects tend to be minimized for the higher energy events because they are usually viewed at large angles relative to the shower axis. It is thus reasonable to assume that the underlying systematic shifts will be most cleanly detected at the higher energies. In this paper, we take the HiRes >$10^{18}$ eV Xmax distribution and the Stereo Fly's Eye >$10^{18}$ eV Xmax distribution and assume that the shift in the mean Xmax, averaged over the energies is representative of the overall systematic shift between these two experiments. We then apply the same systematic shift to the lower energy Stereo Fly's Eye data and check for consistency with the HiRes Prototype/MIA data. The mean Xmax shift is 13 gm/cm$^2$, well within the quoted systematic errors on both experiments. Fig 2 shows the elongation rate data for all three experiments with all the Fly's Eye stereo points shifted upwards by 13 gm/cm$^2$. The data over the entire energy range are quite consistent, even within the statistical errors on the mean. While the HiRes elongation rate above $10^{18}$ eV is consistent with the shifted Stereo Fly's Eye points above this energy, the data can also be fit over the entire energy range to a single elongation rate slope of 68.5 ± 1.5 gm/cm$^2$. Combining all the data after a simple shift consistent with systematic errors shows both the good consistency in the data itself and the importance of a wide dynamic range in energy in interpreting apparent changes in the elongation rate.

## 3. Comparison of Xmax Distributions

The fluctuations of showers in the atmosphere are also an important indication of the cosmic ray composition. From very simple superposition arguments, one expects EAS produced by heavy nuclei such as Fe to have smaller fluctuations and occur higher in the atmosphere than EAS produced by protons. Detailed comparisons to predictions have been made for each of the experiments. Here we examine the consistency of the data for the three experiments.

We ask if the Stereo Fly's Eye stereo, HiRes/MIA and HiRes experiments are consistent in the reported widths of the measured Xmax distributions. For historical reasons, we consider the Xmax distributions in energy bins of $<3\times10^{17}$ eV (HiRes prototype/MIA ), 3 to $5\times10^{17}$ eV (Stereo Fly's Eye and HiRes prototype/MIA), $5\times10^{17}$ to $1\times10^{18}$ eV ( Fly's Eye Stereo) and $>1\times10^{18}$ eV( Stereo Fly's Eye Stereo and HiRes). Fig 2 and Fig 3 show the normalized Xmax distributions for the two energy bins where we have data from two different experiments.

Note that we have shifted the Xmax for all Stereo Fly's Eye events by the canonical 13 gm/cm$^2$, as above. The Xmax distributions are quite consistent. Note that the quoted Xmax resolutions for Stereo Fly's eye and HiRes/MIA are quite similar, with similar resolutions. HiRes has significantly better resolution than Stereo Fly's Eye above $10^{18}$ eV, but the fluctuations are likely dominated by the large intrinsic proton fluctuations and hence the difference in detector resolution is not noticeable. Fig 3 shows the normalized Xmax distributions for all the energy bins in one figure (for clarity we have only used the Stereo Fly's Ey and the HiRes distribution in the 3-$5\times10^{17}$ eV and $>10^{18}$ eV energy bins respectively.) It is clear that the Xmax distributions broaden with increasing energy, as expected for a transition from a heavy to a light composition.

## 4. Conclusion

While additional small systematic shifts from, for example, Čerenkov subtraction, may be present at energies lower than $10^{18}$ eV, a simple 13 gm/cm$^2$ shift of the Stereo Fly's Eye data brings all Xmax distributions and the elongation rate into reasonable agreement. The break in the elongation rate, indicating a transition from a region of changing composition to a light composition appears to lie somewhere between $10^{18}$ and $3\times10^{18}$ eV in the combined data . It will be important to elucidate this with higher precision in a single experiment with a broad dynamic range in energy, since the interpretation of the ankle structure as either a dip due to e+e- pair production of extragalactic protons off the 2.7 degree BB radiation or a transition from a galactic to an extragalactic spectrum very much depends on it[4].

## References

[1] D.J. Bird et. al., Ap. J. 424, 491 (1994); D. J. Bird et. al., Phys. Rev. D 47, 1919 (1993).
[2] T. Abu-Zayyad et. al., Phys. Rev. Lett. 84, 4276 (2000); T. Abu-Zayyad et al., Ap. J. 557, 686(2001).
[3] T. Abu-Zayyad et al., Ap. J. 622, 910 (2005).
[4] V. Berezinsky, A. Z. Gazizov and S. I. Grigorieva, Phys.Lett. B612 (2005) 147-153

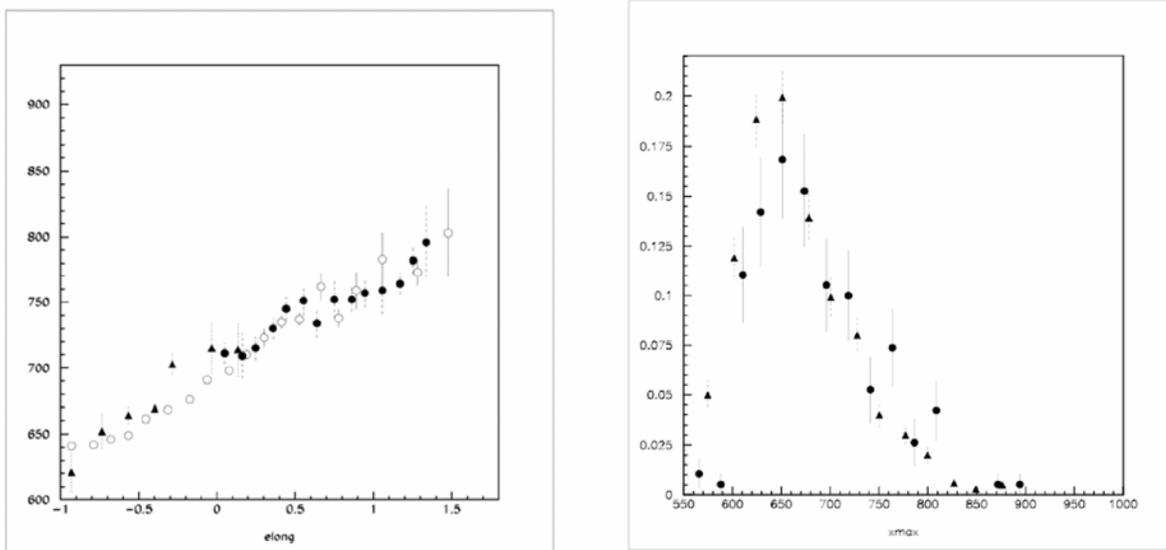

**Fig. 2** . Left: Elongation rate. Open circles – Stereo Fly's Eye; Filled cirles – HiRes ; Triangles – HiRes/MIA. Stereo Fly's Eye data has been shifted by 13 gm/cm². Errors are errors on the mean. Right: Xmax Distribution in $3\times10^{17}$ to $5\times10^{17}$ eV bin. Triangles:Stereo Fly's Eye Stereo; Circles: HiRes/MIA.

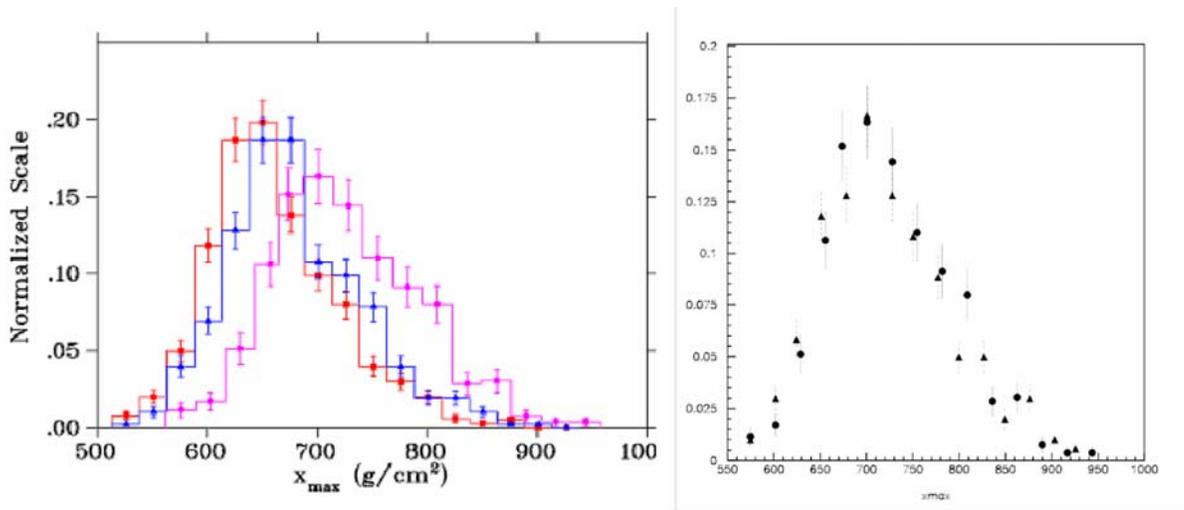

**Fig. 3.** Right: Xmax distribution in $>10^{18}$ eV energy bin. Triangles: Stereo Fly's Eye; Circles: HiRes. Left: Sequence of Xmax distributions in three increasing energy bins $3\text{-}5\times10^{17}$, $5\text{-}10\times10^{17}$ and $> 10^{18}$eV.